\documentclass[aip,jap,reprint,amssymb,amsmath,floatfix,makecell]{revtex4-2}

\usepackage[colorlinks=true,linkcolor=blue,citecolor=blue,urlcolor=blue]{hyperref}
\usepackage{graphicx,bm}

\begin{document}

\title{Competition between isotropic and strongly anisotropic terms in the impact ionization rate of narrow- and middle-gap cubic semiconductors}

\author{A.~N.~Afanasiev}
\email{afanasiev.an@mail.ru} 

\author{A.~A.~Greshnov}

\author{G.~G.~Zegrya}
\affiliation{Ioffe Institute, St.~Petersburg 194021, Russia}

\date{\today}

\begin{abstract}
We report on the strong anisotropy of the inter-band process of impact ionization in direct-gap cubic semiconductors with either weak or strong spin-orbit coupling at low effective temperatures of electron distribution $T$, and the crossover to isotropic behavior with increasing $T$. Such anisotropy is related to specific mechanism of the impact ionization involving coupling of the electron and heavy hole states $\it via$ remote bands, which is vanishing for some high-symmetry propagation directions of an initial electron, namely [100] and [111]. At room temperature impact ionization rate in narrow-gap semiconductors InSb, InAs, GaSb and $\mathrm{In}_{0.53}\mathrm{Ga}_{0.47}\mathrm{As}$ is isotropic while in middle-gap InP, GaAs and CdTe both terms are comparable. We propose simple and justified analytic generalization of Keldysh formula for the impact ionization rate valid for direct-gap semiconductors with $E_g$ up to 1.5 eV, which is suitable for incorporation into modelling software.
\end{abstract}

\maketitle


\section{Introduction\label{sec1}}
\begin{figure}[t!]
 \includegraphics[width=\linewidth]{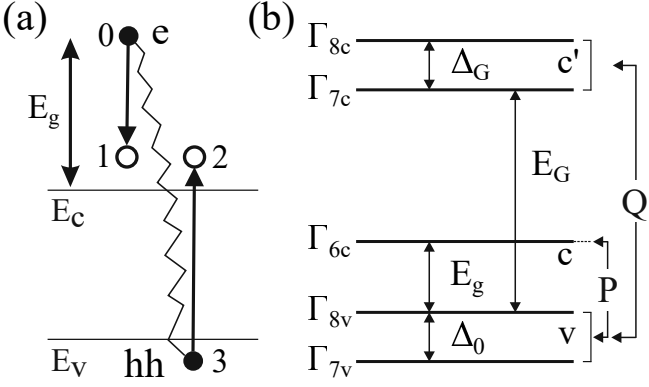}
 	\caption{\label{Fig1} a) Cartoon of the elementary act of impact ionization; b) Schematic representation of the 14-band $\bf{k}\cdot\bf{p}$ model.
	}
\end{figure}

A phenomenon of the interband impact ionization consisting in generation of the electron-hole pairs forced by the Coulomb interaction between a hot conduction band electron and the electrons filling the valence band, as illustrated in Fig.~\ref{Fig1}a, plays an important role in many electronic devices. In some of them, {\it viz.} the conventional semiconductor diodes and MOSFETs, the avalanche breakdown caused by impact ionization restricts the operation voltages, so traditionally the effect is perceived as \emph{negative}. Actually, it is \emph{positive} in many other devices, for which multiplication of carriers due to impact ionization forms a principle of operation. Among them are avalanche transit-time diodes (IMPATT), avalanche photodiodes (APD)~\cite{Sze2021} and a transistor with field-effect control of impact ionization (I-MOS)~\cite{iMOS}, which demonstrates the slope of the subthreshold part of the I-V characteristics up to $5\,\textrm{mV/dec}$ at $T=400\,\textrm{K}$, leading to significant reduction of the switching times in comparison to the conventional devices.

Nowadays, numerical modelling of the physical processes occurring inside the semiconductor devices has become inherent part of engineering, but often the physical models embedded to software are phenomenological and inaccurate, with too many tuning parameters used. Since in practice the output characteristics arising from impact ionization depend on many details, including specific band structure of a given material and characteristics of the relaxation processes giving a particular form of the non-equilibrium distribution function, the problem of calculation, say, of the I-V characteristics, starting directly from a band structure model is in no sense easy, from both conceptual and technical points of view. Therefore, the most popular method for modelling is Monte Carlo~\cite{Trumm2006,Chen2008,Bertazzi2009,Chia2010,Bellotti2012,Shishehchi2013,Asmontas2013,Kodama2013,Ghosh2018,Asmontas2019,Asmontas2020}, but it is obviously dependent on a particular relation between the impact ionization rate and the energy of a hot electron initiating the process, ${\mathcal W}(E)$. Phenomenologically, the rate must grow like a power of the excess energy above a threshold,
\begin{equation}
 	{\mathcal W}(E)=C(E-E_{\mathrm{th}})^n,
 \label{eq0}
\end{equation}
but the values of $n$ and $C$ cannot be revealed without quantum-mechanical calculations, and another question is how far from a threshold this trend holds. While the most popular in literature~\cite{Fischetti1988}, quadratic dependence ($n=2$) was first given by Keldysh~\cite{Keldysh1960} from quite general arguments more than half a century ago, a coefficient before the second power of the excess energy in~(\ref{eq0}) has never been calculated analytically except for an estimation by the ''f-sum'' rule~\cite{Ridley2013}, which turned out to be much larger than a value obtained from numerical calculations using the 30-band $\bf{k}\cdot\bf{p}$ model~\cite{Burt1984}. Some textbooks even advise looking at the prefactor as an adjustable parameter ''to agree with experimental results'' at fixed $n=2$ (see~\cite{Brennan1999}, p.~511). Actually, troubles with quadratic contribution stem from the fact that it vanishes within the spherically symmetric bands, which is the case of the simplest 8-band $\bf{k}\cdot\bf{p}$ model which takes into account the direct coupling between the $s$-type conduction band and the $p$-type valence band only. Therefore, calculation of the Coulomb matrix element entering analytic expression for the impact ionization rate requires going beyond this approximation and taking into account coupling to the remote bands~\cite{Gelmont1978}, and a magnitude of the quadratic term turns out to be rather small. This fact was confirmed by some numerical studies~\cite{Beattie1990}, and rather cubic ($n=3$) than quadratic dependence was found. Analytic expression for the cubic term can be found in the only paper by Gelmont {\it et al.}~\cite{Gelmont1992}, but without any derivation. Thus, a proper analytic answer for ${\mathcal W}(E)$ has been inaccessible to the specialists in Monte Carlo modelling, so they prefer using some arbitrary values of the power $n$ (and prefactor $C$) such as $n=2.5$ and $n=4.3$~\cite{Choo2004}, $n=5.2$~\cite{Trumm2006}, $n=3$~\cite{Chen2008,Chia2010,Chia2013}, $n=3.9$~\cite{Dolgos2012}, $n=1.85$~\cite{Sandall2013}. Some theoretical studies were focused on giving efficient recipes for the proper choice and numerical solution of the band models suitable for the realistic modelling~\cite{Abram1988,Brand1984}, but incorporation of the band calculations into Monte Carlo modelling seems too complicated to be practical. So far, most real engineering calculations of devices just use~(\ref{eq0}) with freely tuned parameters $n$ and $C$, resulting in uncontrollable results.

The aim of this paper is to shed light on real form of ${\mathcal W}(E)$ for the direct-gap semiconductors under practically important conditions implying that the effective temperature of the non-equilibrium distribution of electrons is of the order of a few tens of meV. We provide explicit analytical expressions for the coefficients in the quadratic and cubic terms and give estimation for the crossover temperature $T^*$ at which the carrier generation rates due to both contributions become equal, under assumption of a model isotropic classical distribution of the non-equilibrium electrons. Our results give qualitative explanation and quantitative criteria for strong domination of the cubic term at room temperature in the narrow-gap semiconductors, while in middle-gap semiconductors both terms are comparable.


\section{Theory\label{sec2}}
Usually, the conduction band electrons are treated as the quasi-particles, which do not interact with the electrons filling the valence bands, and even the process of (chcc-type) Auger recombination can be viewed as a result of the interaction of just two electrons, one going to a higher state in the conduction band and another one to a free state in the valence band. For the interband impact ionization this is not true since it is initiated by the Coulomb interaction between a hot conduction band electron and all the electrons of the valence bands, the fact easily verified within the Hartree-Fock approximation. Therefore, a partial rate due to the elementary process sketched in Fig.~\ref{Fig1}a, given by
\begin{equation}
 	W = \frac{2\pi}{\hbar}
 	\left| \left\langle \alpha_1\alpha_2
 	\left| \frac{e^2}{\varkappa |{\bf r}_1-{\bf r}_2|} \right|
 	\alpha_{0}\alpha_{3} \right \rangle \right|^2
 	\delta(\Delta E),
 \label{Prob_comm}
\end{equation}
must be summed over the possible initial states of the electron 3 in the valence bands and the possible final states of the electrons 1 and 2 in the conduction band to obtain the total impact ionization rate due to a given hot electron ''0''. Here $\Delta E = E_1+E_2-E_0-E_3$ is the energy balance and $\alpha_{i}=\{{\bf k}_i, \xi_i\}$ denotes a full set of the quantum numbers -- wave vector $\textbf{k}_i$ and total angular momentum projection $\xi_i$ -- in the conduction (i=0, 1 or 2) or the valence band (i=3). Concerning the latter, we consider only the processes involving the heavy hole states since the light hole and spin-orbit split hole states lie well below in energy for the wave vectors bigger than the threshold one and are relevant for the very hot and rare initial electrons only. Applying the Fourier transform and integrating out the zero-$\bf k$ Bloch functions, it is straightforward~\cite{Ridley2013,Brennan1999} to rewrite~(\ref{Prob_comm}) as
\begin{equation}
\label{Prob_final}
 	W=\frac{2\pi}{\hbar} \left(\frac{4 \pi e^2}{\varkappa}\right)^2
 	\frac{I_{cc}(\alpha_0,\alpha_1) I_{cv}(\alpha_2,\alpha_3)}{|{\bf k}_0-{\bf k}_1|^4}
 	\delta_{\Delta{\bf k},0} \delta(\Delta E),
\end{equation}
where the squared overlap integrals of the Bloch functions $I_{cc}$ and $I_{cv}$ can be expressed in terms of the state vectors $|{\mathcal F}\rangle$ defined in a basis of the $\Gamma$-point Bloch functions $u^{(0)}_n({\bf r})$ as $I_{c,c/v}(\alpha_i,\alpha_j)=|\langle{\mathcal F}_{\alpha_i}|{\mathcal F}_{\alpha_j}\rangle|^2$ and $\Delta{\bf k}={\bf k}_0+{\bf k}_3-{\bf k}_1-{\bf k}_2$ denotes the momentum balance. The energy and momentum conservation laws expressed by the delta functions in~(\ref{Prob_final}) impose restrictions on the wave vectors ${\bf k}_i$, which set the following impact ionization threshold when non-parabolicity in dispersion of the initial electron is taken into account at $\mu=m_e/m_{hh}\ll1$~\cite{Volkov1976}:
\begin{gather}
 {\bf k}_{0}^{\mathrm{th}} = {\bf k}_g (1+3\mu/2), \label{thres_k0} \\
 E_{\mathrm{th}} = E_e({\bf k}_{0}^{\mathrm{th}}) = E_g (1+2\mu), \label{thres_en} \\
 {\bf k}_{3}^{\mathrm{th}} = -{\bf k}_g (1-\mu/2), \label{thres_k3} \\
 {\bf k}_{1}^{\mathrm{th}} = {\bf k}_{2}^{\mathrm{th}} = \mu {\bf k}_g , \label{thres_k12}
\end{gather}
where $k_g=\frac{2}{\hbar}\sqrt{F_1(\Delta_0/Eg)m_e E_g}$ with $F_1(x) = \frac{(1+2x/3)(1+x/2)}{(1+x)(1+x/3)}$. Since $\bf{k}\cdot\bf{p}$ coupling between the conduction band $\Gamma_{6c}$ (or "$c$" in Fig.~\ref{Fig1}b) and valence bands $\Gamma_{8v}$ and $\Gamma_{7v}$ (or ''$v$'' in Fig.~\ref{Fig1}b) does not contribute to dispersion of the heavy holes, the smallness of $\mu$ is equivalent to $E_g/E_G \ll 1$, where $E_G$ is a minimal distance from $v$ to the bands contributing to the inverse heavy hole mass ($c'$ band in the 14-band $\bf{k}\cdot\bf{p}$ model used in this work, see Fig.~\ref{Fig1}b). Spin-orbit splittings of $c'$ and $v$ band are also small compared to the $c'-v$ distance $\Delta_{0,G}/E_G \ll 1$~\cite{Winkler2003} (see Table~\ref{Tab:Band_Parameters}).

\begin{table*}[t!]
	\caption{\label{Tab:Band_Parameters} Band structure parameters (taken from~\cite{Winkler2003,Levinshtein1996,Cardona1988,NSMbase,Fonthal2000,Vurgaftman2001}) of narrow- and middle-gap semiconductors within the 14-band $\textbf{k}\cdot\textbf{p}$-model used in this work and the corresdonding values of the dimensionless parameters $\beta$~(\ref{eq_critso}), $x=\Delta_0/E_g$ and the crossover effective temperature~(\ref{Teff}).}
	\begin{ruledtabular}
		\begin{tabular}{ccccccccc}
			& $E_g$ (eV)\footnotemark[1] & $\Delta_0$ (eV) & $P$ ($\mathrm{eV}\cdot \mathrm{\AA}$) & $E_{G}$ (eV) & $Q$ ($\mathrm{eV}\cdot \mathrm{\AA}$) & $\beta$\footnotemark[2] & $x$\footnotemark[2] & $T^*$ (K)\footnotemark[3]\\
			\hline
			InSb & 0.24 & 0.81 & 9.64 & 3.2 & 8.13 & 18.78 & 4.5 & 3.4 \\
			InAs & 0.42 & 0.39 & 9.2 & 4.4 & 8.33 & 2.72 & 1.09 & 13.4 \\
			GaSb & 0.81 & 0.76 & 9.62 & 3.3 & 8.11 & 1.12 & 1.05 & 121 \\
			$\mathrm{In}_{0.53}\mathrm{Ga}_{0.47}\mathrm{As}$ & 0.82 & 0.33 & 9.81 & 4.4 & 8.25 & 0.62 & 0.44 & 76 \\
			InP & 1.42 & 0.11 & 8.85 & 4.7 & 7.22 & 0.07 & 0.08 & 304 \\
			$\mathrm{In}_{0.52}\mathrm{Al}_{0.48}\mathrm{As}$ & 1.53 & 0.3 & 9.09 & 4.5 & 8.25 & 0.13 & 0.21 & 507\\
			GaAs & 1.52 & 0.34 & 10.49 & 4.5 & 8.17 & 0.21 & 0.24 & 306 \\
			CdTe & 1.61 & 0.95 & 9.5 & 5.4 & 7.87 & 0.54 & 0.63 & 310\\
		\end{tabular}
	\end{ruledtabular}
\footnotetext[1]{At $T=0$ K.}
\footnotetext[2]{At $T=296$ K.}
\footnotetext[3]{Calculated using the temperature-dependent values of bandgaps from~\cite{NSMbase,Fonthal2000,Vurgaftman2001}.}
\end{table*}

In practice, distribution function of the ''impact ionization ready'' electrons extends on much smaller scale (say, $25\,\textrm{meV}$) than the value of $E_{\mathrm{th}}$, therefore it is convenient to introduce the ''above-threshold'' components of the wave vectors according to ${\bf q}_i={\bf k}_i-{\bf k}_i^{\mathrm{th}}$, assuming that $(E-E_{\mathrm{th}})/E_{\mathrm{th}} \ll 1$. Under such approximations the rate of impact ionization due to electron in a state $\alpha_0$ reduces to
\begin{multline}
	{\mathcal W} =\frac{\pi\hbar F_2\left(\frac{\Delta_0}{Eg}\right)}{12m_e E_g^2} \left(\frac{4\pi e^2}{\varkappa}\right)^2 \int\frac{d^3q_1d^3q_2}{(2\pi)^6} [\widetilde{I}_{cv}({\bf q}_1,{\bf q}_3)\\
	+ \widetilde{I}_{cv}({\bf q}_2,{\bf q}_3)] \delta\left(q^2_1+q^2_2-\frac{2m_e(E_0-E_{\mathrm{th}})}{\hbar^2}\right),
 \label{rate_final}
\end{multline}
where 
\begin{multline}
	\widetilde{I}_{cv}({\bf q}_i,{\bf q}_3) =\\
	 \sum_{\xi_i,\xi_3} I_{cv}({\bf k}_{i}^{\mathrm{th}}+{\bf q}_i,\xi_i;{\bf k}_{3}^{\mathrm{th}}+{\bf q}_1+{\bf q}_2-{\bf q}_0,\xi_3),
\end{multline}
is the interband overlap integral summed over the total angular momentum projections (on the direction of ${\bf k}_3$) of heavy hole $\xi_3 = \pm \frac{3}{2}$ and final electron $\xi_i = \pm \frac{1}{2}$ states, while
\begin{equation}
 	F_2(x)=\frac{(1+x)^2\left(1+\frac{x}{3}\right)^3} {\left(1+\frac{7}{9}x+\frac{x^2}{6}\right)\left(1+\frac{2}{3}x\right)^2\left(1+\frac{x}{2}\right)}
\end{equation}
is equal to 1 for both $\Delta_0\ll E_g$ and $\Delta_0\gg E_g$ and the excess wave vector of the initial electron above the threshold one $q_0=\left(\frac{\partial E_0}{\partial k_0}\right)_{\mathrm{th}}^{-1}(E_0-E_{\mathrm{th}})$ is assumed to be collinear to ${\bf k}^{\mathrm{th}}_0$. Expression~(\ref{rate_final}) shows that the energy (and, actually, angular) dependence of the impact ionization rate ${\mathcal W}$ is governed by dispersion of the squared overlap integral $\widetilde{I}_{cv}$ near ${\bf q}_{1,2}=0$. Since $\widetilde{I}_{cv}$ expresses a degree at which the states of the conduction and valence bands are overlapping, in practice it is strongly dependent of a particular considered multi-band model. A minimal basis for such model consists of the two $s$-type and six $p$-type $\Gamma$-point Bloch functions $u^{(0)}_n({\bf r})$, and a minimal coupling is direct $\bf{k}\cdot\bf{p}$ coupling between the $s$-type and $p$-type states, described by the only matrix element $P$~\cite{Kane1956}, which is also known as the Kane matrix element. This 8-band model nicely describes dispersion of the electron and light holes in the narrow-gap semiconductors, but not the heavy holes, which remain dispersionless. Near a threshold $k_i\ll k_3$ ($i=1,2$) and explicit expression for the squared overlap integral in this approximation is given by
\begin{equation}
 	I_{cv}(\alpha_i,\alpha_3) = \frac{P^2|[{\bf k}_i\times{\bf k}_3]|^2}{2E_g^2k_3^2} \delta_{|\xi_i-\xi_3|,1}.
 \label{eq_Icv8k}
\end{equation}
The property of $I_{cv}$ for vanishing at the collinear ${\bf k}_i$ and ${\bf k}_3$ is more general and applicable to the threshold values of the wave vectors given by Eqs.~(\ref{thres_k3}),(\ref{thres_k12}). Therefore $\widetilde{I}_{cv}$ can be approximated by
\begin{equation}
 	\widetilde{I}_{cv}({\bf q}_i,{\bf q}_3) = \frac{P^2 q_{i\perp}^2}{E_g^2} = \frac{\hbar^2q_{i\perp}^2}{2m_e} \frac{1+\frac{\Delta_0}{E_g}}{E_g+\frac{2}{3}\Delta_0},
 \label{eq_Icv8q}
\end{equation}
where ${\bf q}_{i\perp}$ is a component of ${\bf q}_{1,2}$ in a plane perpendicular to wave vector of the initial electron. Straightforward integration of $\widetilde{I}_{cv}$ according to~(\ref{rate_final}) leads to the following expression for $\mathcal W$, which is cubic in $E-E_{\mathrm{th}}$:
\begin{gather}
 	\mathcal{W}_3(E) = B(E-E_{\mathrm{th}})^3, \\
 	B= \frac{\omega_B^*}{18 E_g^3} \frac{E_g+\Delta_0}{E_g+\frac{2}{3}\Delta_0} F_2\left(\frac{\Delta_0}{E_g}\right),
 \label{W3}
\end{gather}
where $\omega_B^*=\frac{m_e e^4}{2 \hbar^3 \varkappa^2}$ is the Bohr frequency for the conduction band electrons. In the limiting case of the infinite spin-orbit splitting $\Delta_0$ (corresponding to the 6-band $\bf{k}\cdot\bf{p}$ model) this answer reduces to the one given implicitly by~(2) of~\cite{Gelmont1992}, while taking into account a finite value of $\Delta_0$ gives an additional factor to the latter, which goes to $2/3$ when $\Delta_0\rightarrow0$.

\begin{figure*}[t!]
 \includegraphics[width=\linewidth]{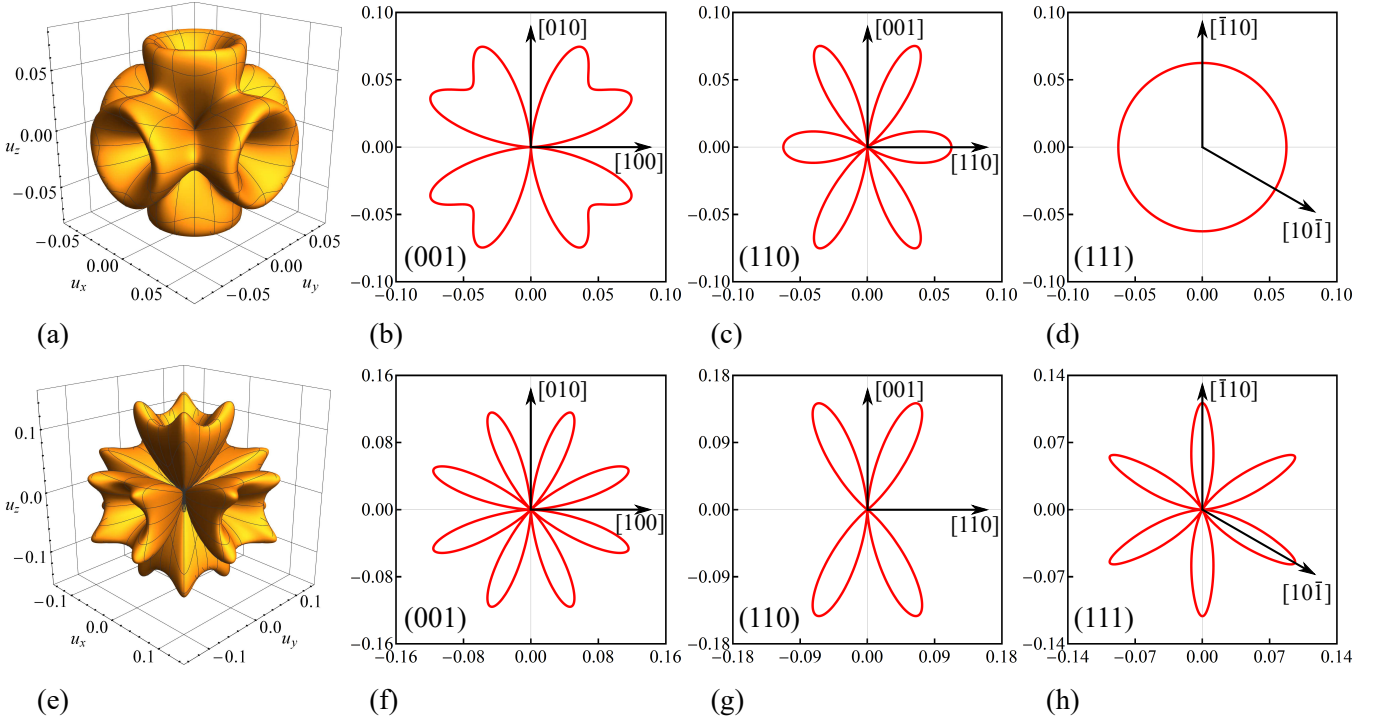}
 	\caption{
		\label{Fig2}
 		Angular plots of $\mathcal{K}(\textbf{u},\beta)$ representing the anisotropy of the quadratic term~(\ref{W2}) in the impact ionization rate for the cases of a)-d) strong $\beta \gg 1$~(\ref{Kinf}) and e)-h) weak $\beta\ll 1$~(\ref{Kzero}) spin-orbit coupling; a) and e) all wave vector directions of the initial electron, $u_{x,y,z}$ denote the projections of $\textbf{u}$ onto the [100] set of equivalent crystallographic directions; b) and f) cross sections by (001) plane; c) and g) cross sections by (110) plane; d) and h) cross sections by (111) plane.
	}
\end{figure*}

Thus, the minimal band model misses a quadratic contribution associated with the value of the interband overlap integral at threshold $\widetilde{I}_{cv}(0,0)$, which arises in more sophisticated models taking into account coupling to the remote bands and lowering the spherical symmetry to the cubic one ($\mathcal{O}_h/\mathcal{T}_d$), in particular, 14-band $\bf{k}\cdot\bf{p}$ model (so called extended Kane model~\cite{Winkler2003}). In this model six additional Bloch states of $\Gamma_{7c}$ and $\Gamma_{8c}$ symmetry lying a few electron-volts above $E_c$ (the second conduction band, $c'$ in Fig.~\ref{Fig1}b) are coupled to the states of the valence band {\it via} the only matrix element $Q$, which is of the same order of magnitude as $P$ (see Table~\ref{Tab:Band_Parameters}). Inversion asymmetry induces coupling between the bands $c$ and $c'$ described by the matrix elements $P'$ and $\Delta'$, which are about an order of magnitude smaller than $P$, $Q$ and $\Delta_0$, correspondingly~\cite{Richard2004}.

In order to treat coupling between $c'$ and $v$ bands perturbatively, it is convenient to divide the full $\bf{k}\cdot\bf{p}$ hamiltonian into the main part ${\mathcal H}_0({\bf k})$, representing hamiltonian of the minimal 8-band model and the energies of the $c'$ band states at $k=0$, and a perturbation ${\mathcal V}({\bf k})$, describing the above-mentioned $c'-v$ coupling. The six eigenstates of ${\mathcal H}_0$ at ${\bf k} = {\bf k}_3^{\mathrm{th}} \approx - {\bf k}_g$ corresponding to $c'$ band lie far away from the rest eight, {\it viz.} the electron states with $E_e=E_v+2E_g$, the heavy hole states with $E_{hh}^{(0)}=E_v$, and the light and spin-orbit split hole states with
\begin{equation}
 	E_{lh/so} = E_v - \frac{E_g}{2} \left(1+x \pm \sqrt{\frac{x^3+x^2-x+3}{x+3}}\right) ,
 \label{eq_lhso}
\end{equation}
where $x=\Delta_0/E_g$. Also, expression~(\ref{eq_lhso}) guarantees that minimal distance in energy between the heavy holes and the other hole branches is bigger than $\min(E_g,\Delta_0)/2$ for $x \geq 1$, which is the case of narrow-gap semiconductors (see Table~\ref{Tab:Band_Parameters}). Consequently, the unperturbed heavy hole state is non-degenerate and the corresponding perturbation theory can be applied. The particular method we follow to calculate the Bloch functions is described in Appendix~\ref{Sec_Ap:Perturbation}. However, the proper parameter which should not be small for the perturbative treatment in the form given in Appendix~\ref{Sec_Ap:Perturbation} to be correct is
\begin{equation}
 	\beta = \frac{P^2 \Delta_0 E_G}{6 Q^2 E_g^2}\gg 1.
 \label{eq_critso}
\end{equation}
Indeed, the formulae~(\ref{eq_hh1})-(\ref{eq_hh2}) show that the $\bf{k}\cdot\bf{p}$ perturbation ${\mathcal V}(-{\bf k}_g)$ is applied to a ''bare'' heavy hole state $|{\mathcal F}_{hh}^{(0)}\rangle$ twice, producing a factor $\sim (Qk_g)^2 \sim E_g^2$, which is compensated by the two Green functions, {\it i.e.} the energy denominators. And while the first denominator is a distance between the bands $c'$ and $v$ given by $E_G$, the second one counts a distance between the heavy holes and either the electrons, light holes or the spin-orbit split holes. When $\Delta_0$ is much smaller than $E_g$ the spin-orbit holes behave more like the heavy holes, and the energy separation between them at finite wave vector $\textbf{k}=\textbf{k}_3^{\mathrm{th}}\approx -\textbf{k}_g$ given by $E_v-E_{so}$ after Eq.~(\ref{eq_lhso}) tends to its value at $k=0$, $\Delta_0$. Thus, it could be deduced that the described perturbative approach does not work at $\beta\ll1$ and it leads to divergent result at $\Delta_0\rightarrow0$. However, while the first is true, the second is not because the unperturbed spin-orbit split hole states do not overlap with the $s$-type states in this limit, transforming into the second branch of the heavy holes. Therefore, for the case of middle-gap semiconductors, when $x=\Delta_0/E_g \ll 1$ (or alternatively $\beta \ll 1$) becomes another small parameter (see Table~\ref{Tab:Band_Parameters}) in addition to $m_e/m_{hh}\ll 1$, our previous method should be rearranged to take into account the degeneracy of heavy holes, {\it viz.} $|\mathcal{F}^{(0)}_{hh}\rangle$ in Appendix~\ref{Sec_Ap:Perturbation} now stands for the proper zero-order wave function, corresponding to the topmost branch of heavy hole states split by $\bf{k}\cdot\bf{p}$ interaction between $c$ and $v$ bands.

\begin{figure}[t!]
 \includegraphics[width=\linewidth]{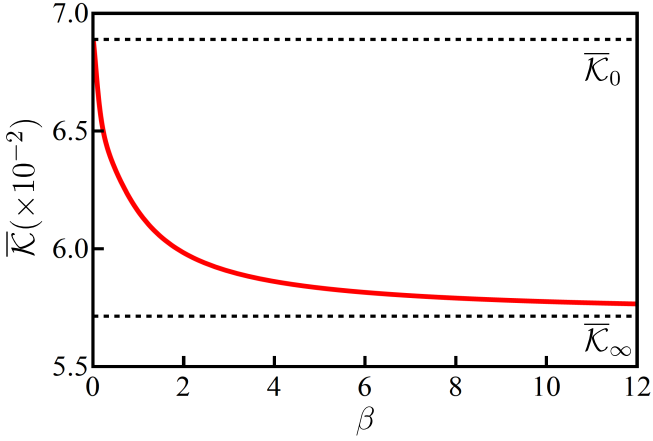}
 	\caption{\label{Fig3} Average value of the cubic invariant approximated by analytical expressions~(\ref{Kapprox})-(\ref{KapproxLast}) for arbitrary parameter $\beta$~(\ref{eq_critso}).
	}
\end{figure}

After squaring and summing over the $\xi$-variables, we obtain explicit expression for the main ingredient of the quadratic contribution to the impact ionization rate,
\begin{equation}
	\widetilde{I}_{cv}(0,0) =\frac{8 E_g^2 }{E^2_G} \frac{Q^4}{P^4} \mathcal{K}({\bf u},\beta)\frac{1+x/2}{1+x/3}
\label{I2}
\end{equation}
where $\mathcal{K}({\bf u},\beta)$ is a cubic invariant, which can be expressed in terms of parameter $\beta$ and the polynomial invariants $\mathcal{I}({\bf u})$ and $\mathcal{J}({\bf u})$ of the fourth and sixth orders
\begin{gather}
	\mathcal{I}({\bf u})= u_x^2u_y^2 + u_x^2u_z^2 + u_y^2u_z^2,\\
	\mathcal{J}({\bf u})= u_x^2u_y^2u_z^2.
\end{gather}
Here ${\bf u} = {\bf k}_0/k_0$ characterizes direction of the initial electron motion with respect to crystallographic axes. For big ($\beta\to \infty$) and small ($\beta \to 0$) values of $\beta$, anisotropy of the quadratic contribution is described by
\begin{equation}
	\mathcal{K}_{\infty}({\bf u}) = \mathcal{I}(1-3\mathcal{I}),
	\label{Kinf}
\end{equation}
and
\begin{multline}
 	\mathcal{K}_0({\bf u}) = \mathcal{K}_{\infty}({\bf u})- \mathcal{I}^2 \\
	+ 3\mathcal{J}+\frac{\mathcal{I}^2(1-4\mathcal{I})-\mathcal{J}(2-9\mathcal{I})}{\sqrt{\mathcal{I}^2-3\mathcal{J}}},
	\label{Kzero}
\end{multline}
respectively. Approximate form of $\mathcal{K}(\textbf{u},\beta)$ for arbitrary $\beta$ is given in Appendix~\ref{Sec_Ap:K_full}. Earlier, a similar expression for the case $\beta\gg1$ was given in~\cite{Gelmont1978} in terms of the Luttinger parameters $\gamma_2$ and $\gamma_3$ for the limiting case of $\Delta_0\rightarrow\infty$ with application to the problem of Auger recombination and in~\cite{Afanasiev2017}. Substituting~(\ref{I2}) into~(\ref{rate_final}) for $\widetilde{I}_{cv}({\bf q}_i,{\bf q}_3)$, we obtain explicit expression for the quadratic term in the impact ionization rate, namely
\begin{gather}
 	\mathcal{W}_2(E,{\bf u}) = A(E-E_{\mathrm{th}})^2,\\
 	A=\frac43 \frac{\omega_B^*}{E_G^2} \frac{Q^4}{P^4} \mathcal{K}({\bf u},\beta) \frac{E_g+\frac12\Delta_0}{E_g+\frac13\Delta_0} F_2\left(\frac{\Delta_0}{E_g}\right).
 \label{W2}
\end{gather}

\begin{figure}[t!]
 \includegraphics[width=\linewidth]{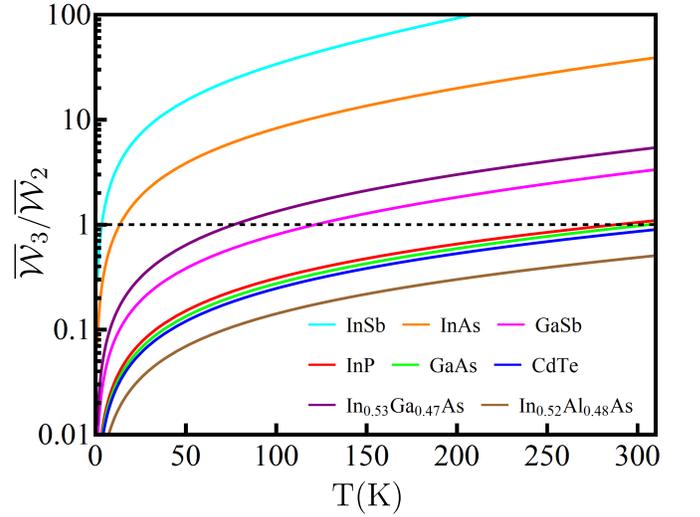}
 	\caption{\label{Fig4} Competition between averaged cubic and quadratic contributions to the impact ionization rate of semiconductors listed in Table~\ref{Tab:Band_Parameters} at various effective temperatures of hot electron distribution above the impact ionization threshold. Intersections of solid lines with dashed one correspond to crossover effective temperatures.
	}
\end{figure}

With $E_G^2$ in the denominator, the quadratic contribution given by~(\ref{W2}) turns out to be of the second order in small parameter $\mu=m_e/m_{hh}$, leading to competition with the cubic contribution given by~(\ref{W3}) in narrow- and middle-gap semiconductors for the electrons with $E-E_{\mathrm{th}}$ of the order of a few tens of $\textrm{meV}$s, as shown below. As illustrated in Fig.~\ref{Fig2}, $\mathcal{W}_2$ strongly depends on the orientation of the initial electron wave vector with respect to crystallographic directions. In both cases of strong $\beta \gg 1$ (see Figs.~\ref{Fig2}a-d) and weak $\beta \ll 1$ (see Figs.~\ref{Fig2}e-h) spin-orbit coupling, quadratic term vanishes along the high-symmetry directions [100] and [111]. However, the anisotropy of $\mathcal{W}_2$ described by $\mathcal{K}_{\infty}(\textbf{u})$ and $\mathcal{K}_{0}(\textbf{u})$ is different: in the latter case the quadratic term additionally vanishes in the [110] direction. In the (111) crystallographic plane quadratic term becomes isotropic for the case of strong spin-orbit coupling (see Fig.~\ref{Fig2}d), since $\mathcal{I}(\textbf{u}_{(111)})=1/4$, while $\mathcal{K}_0(\textbf{u}_{(111)})$ reproduces the nontrivial angular dependence of $\mathcal{J}(\textbf{u})$ (see Fig.~\ref{Fig2}h). Interestingly, the spin-orbit interaction and inversion asymmetry lead to some specific contribution to ${\mathcal W}_2$ in the semiconductors belonging to ${\mathcal T}_d$ symmetry group, non-vanishing along the primary crystallographic directions. However, this contribution is small in parameter $\Delta_{c'v}/\Delta_0$, where $\Delta_{c'v}$ is a magnitude of the non-diagonal spin-orbit $c'-v$ coupling~\cite{Winkler2003}, therefore it is very small from a practical point of view.

\begin{figure}[t!]
 \includegraphics[width=\linewidth]{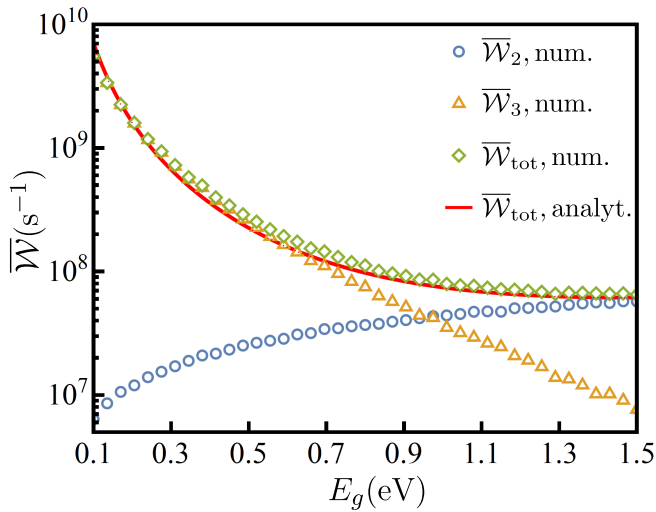}
 	\caption{\label{Fig5} Dependence of the averaged impact ionization rates at $T=296 K$ on the bandgap. Solid red line corresponds to the total rate $\overline{\mathcal{W}}_{\mathrm{tot}}=\overline{\mathcal{W}}_2+\overline{\mathcal{W}}_3$ deduced from analytical expressions~(\ref{W3}) and~(\ref{W2}); open markers denote the behavior of the numerically calculated total rate (green diamonds) and partial contributions to it: quadratic (blue circles) and cubic (orange triangles) ones.
	}
\end{figure}


\section{Discussion\label{sec3}}


To compare the relative importance of the two contributions to total impact ionization rate
\begin{equation}
	\label{Wall}
	\mathcal{W}_{\mathrm{tot}}(E,\textbf{u})=A(\textbf{u}) (E-E_{\mathrm{th}})^2+B(E-E_{\mathrm{th}})^3 ,
\end{equation}
we consider  a non-degenerate ensemble of electrons driven from equilibrium by electric field and calculate the carrier generation rates $\mathcal{R}_2$ and $\mathcal{R}_3$ (corresponding to the scattering rates given by Eqs.~(\ref{W2}) and~(\ref{W3}), respectively), averaged over the field direction. Since the problem of such averaging is equivalent to averaging over directions of the initiating electrons $\bf u$ under the assumption of isotropic distribution, the total rates can be written as
\begin{gather}
	\mathcal{R}_i=\overline{\mathcal{W}}_i \mathcal{N}_0, \\
	\overline{\mathcal{W}}_i=\int\limits_{E_{\mathrm{th}}}^{+\infty}\frac{d E}{T}\frac{d\textbf{u}}{4\pi}\mathcal{W}(E,\textbf{u}) \exp\left(-\frac{E-E_{\mathrm{th}}}{T}\right),
	 \label{eq_Ri}
\end{gather}
where $\mathcal{N}_0=\mathcal{D}(E_{\mathrm{th}})\overline{\delta f}(E_{\mathrm{th}}) T$ is the nonequlibrium concentration of hot electrons above the impact ionization threshold, $\overline{\mathcal{W}}_i$ is the impact ionization rate, averaged over momentum direction of initial electrons and their distrubution, $\mathcal{D}(E)$ is the density of states in the conduction band, $\delta f (E)$ is the non-equilibrium component of the distribution function, $T$ is an effective temperature, determined by either the external temperature, the energy acquired by a mean free path $eEl$, or its combination with the optical phonon energy $\hbar\omega_o$~\cite{Ridley2013}. Performing the elementary integration of $(E-E_{\mathrm{th}})^n$ with the exponential function, we obtain the crossover temperature at which $\overline{\mathcal{W}}_2=\overline{\mathcal{W}}_3$ (or $\mathcal{R}_2=\mathcal{R}_3$):
\begin{equation}
 	T^*=\frac{\overline{A}}{3B}=8\frac{Q^4}{P^4}\frac{E_g^3}{E^2_G}\overline{\mathcal{K}}(\beta)F_1\left(\frac{\Delta_0}{E_g}\right).
 \label{Teff}
\end{equation}
Here the averaged (over the directions of $\textbf{u}$) value of the cubic invariant at arbitrary $\beta$ is between its limits at infinite and zero spin-orbit coupling $\overline{\mathcal{K}}_{\infty}<\overline{\mathcal{K}}(\beta)<\overline{\mathcal{K}}_0$, with $\overline{\mathcal{K}}_{\infty}=0.057$ and $\overline{\mathcal{K}}_0=0.069$. In this work we use the $\overline{\mathcal{K}}(\beta)$ dependence (see Fig.~\ref{Fig3}) deduced from the analytical approximation to $\mathcal{K}(\textbf{u},\beta)$ by Eqs.~(\ref{Kapprox})-(\ref{KapproxLast}). The definition of $F_1(x)$ is given after~(\ref{thres_k12}).

Temperature dependence of the bandgap $E_g(T)$ leads to non-linear scaling of $\overline{\mathcal{W}}_3/\overline{\mathcal{W}}_2$ ratio with $T$, and Eq.~(\ref{Teff}) becomes transcendental. Using band structure parameters listed in Table~\ref{Tab:Band_Parameters} and empirical temperature dependencies of the bandgaps (namely, Manoogian–Wooley equation for CdTe~\cite{Fonthal2000} and Varshni equation for other compounds~\cite{NSMbase,Vurgaftman2001}), we calculate the $\overline{\mathcal{W}}_3(T)/\overline{\mathcal{W}}_2(T)$ dependencies for narrow- and middle-gap semiconductors (see Fig.~\ref{Fig4}) and estimate the crossover temperatures, see Table~\ref{Tab:Band_Parameters}. At low effective temperature, the impact ionization rate of any semiconductor is determined by quadratic term, and thus is strongly anisotropic. With increasing $T$, the isotropic cubic term rapidly grows and at room temperature it completely dominates over $\mathcal{W}_2(E,\textbf{u})$ in narrow-gap semiconductors like InSb, InAs, GaSb and $\mathrm{In}_{0.53}\mathrm{Ga}_{0.47}\mathrm{As}$, while in middle-gap InP, GaAs and CdTe both terms are comparable. In $\mathrm{In}_{0.42}\mathrm{Al}_{0.58}\mathrm{As}$ the crossover takes place at much higher temperature about 500~K.


\section{Conclusion\label{sec4}}

In conclusion, we would like to shortly discuss the accuracy of the obtained analytical expressions. Since the main complexity and therefore potential inaccuracy is concentrated in the proper expression for overlap integrals, determined by the specific form of multiband wave functions, we performed numerical calculation of the averaged impact ionization rate based on the numerical diagonalization of the 14-band \textbf{kp}-model. In order to reduce the numerical complexity, in~(\ref{rate_final}) we have ignored the inessential dependence of the numerically calculated overlap integrals on $\textbf{q}_3$ and preformed the analytical integration of the total rate over magnitudes $q_1$ and $q_2$ and solid angles $\Omega_0$. As a result, 9-dimensional integration over $\textbf{q}_{0,1,2}$ was reduced to 5-dimensional integration over magnitude $q_0$ and solid angles $\Omega_{1,2}$. Finally, the infinite integration interval over $q_0$ was rearranged into [0,1] by means of Lambert function substitution and the resulting integral was calculated by Monte-Carlo method. The obtained results of the averaged impact ionization rate dependence on $E_g$ at the effective temperature $T=296 K$ are presented on the Fig.~\ref{Fig5}. Simple analytical expression $\overline{\mathcal{W}}_{\mathrm{tot}}=2\overline{A}T^2 +6 B T^3$ for the impact ionization rate deduced from Eqs.~(\ref{Wall}) and~(\ref{eq_Ri}) is in good agreement with numerical results in the wide range of $E_g$ up to 1.5 eV. Namely, the corresponding mean percentage error is 11\% and the maximal discrepancy with respect to numerical calculations is 15\%. The values of analytical and numerical "crossover bandgaps" [when $\overline{\mathcal{W}}_2(E_g)=\overline{\mathcal{W}}_3(E_g)$] are also close: 1.15 eV vs 0.95 eV, respectively.

Discrepancy between analytical and numerical results originates from Eq.~(\ref{W2}), which underestimates the quadratic term, especially for the case of wide-gap semiconductors, when the primary small parameter of our theory $\mu=m_e/m_{hh}$ approaches unity. However, 5\% agreement with numerical results within the full range of bandgaps considered in this work can be achieved using analytical expressions, which include higher-order corrections to~(\ref{W3}) and~(\ref{W2}) by $E_g/E_G$ (up to second order). We also expect, that in the case of strongly anisotropic distribution of hot electrons in high electric fields~\cite{Dmitriev1982}, the angular dependence of carrier generation rate will replicate the anisotropy of the total impact ionization rate~(\ref{Wall}). Therefore, the obtained generalization of the conventional Keldysh formula for the impact ionization rate in direct-gap semiconductors given by~(\ref{Wall}) is suitable for incorporation into modelling software.




\section*{Conflict of Interest}

The authors have no conflicts to disclose.


\section*{Data Availability}

The data that support the findings of this study are available from the corresponding author upon reasonable request.

\appendix

\section{Perturbation method description}
\label{Sec_Ap:Perturbation}

To describe the perturbative calculation of the multi-band wave function of the heavy hole state, it is convenient to introduce the ''non-interacting'' Green function ${\mathcal G}_0({\mathcal E})=({\mathcal E}-{\mathcal H}_0)^{-1}$. The first-order correction to the heavy hole eigenstate $|{\mathcal F}_{hh}^{(0)}\rangle$ is then expressed in the form
\begin{equation}
 	|{\mathcal F}_{hh}^{(1)}\rangle=\lim_{{\mathcal E} \to \widetilde{E}_v} {\mathcal G}_0({\mathcal E},-{\bf k}_g) {\mathcal V}(-{\bf k}_g) |{\mathcal F}_{hh}^{(0)}\rangle.
 \label{eq_hh1}
\end{equation}
Due to definition of ${\mathcal V}$ and ${\mathcal G}_0$ in Section~\ref{sec2}, $|{\mathcal F}_{hh}^{(1)}\rangle$ belongs to the $c'$ subspace and is orthogonal to the eight basis states of $c$ and $v$ bands, therefore it does not contribute to $\widetilde{I}_{cv}(0,0)$ as well as to the first-order correction to energy, $E_{hh}^{(1)}=0$. The second-order correction to the energy of the heavy hole state $E_{hh}$ is
\begin{equation}
 	E_{hh}^{(2)}=\lim_{{\mathcal E} \rightarrow \widetilde{E}_v} \langle {\mathcal F}_{hh}^{(0)}| {\mathcal V}(-{\bf k}_g) |{\mathcal F}_{hh}^{(1)}({\mathcal E})\rangle,
 \label{eq_hhen2}
\end{equation}
and the corresponding multiband wave function $|{\mathcal F}_{hh}\rangle$ can be written as
 \begin{multline}
 	|{\mathcal F}_{hh}^{(2)}\rangle= \lim_{{\mathcal E} \rightarrow \widetilde{E}_v} {\mathcal G}_0({\mathcal E},-{\bf k}_g)\\
 		\times\left[{\mathcal V}(-{\bf k}_g)|{\mathcal F}_{hh}^{(1)}({\mathcal E})\rangle-E_2({\mathcal E})|{\mathcal F}_{hh}^{(0)}\rangle \right].
 \label{eq_hh2}
\end{multline}
Equation~(\ref{eq_hhen2}) specifies the heavy hole energy and relation between the heavy hole mass and the 14-band model parameters $Q$ and $E_G$, while Eq.~(\ref{eq_hh2}) gives principal approximation for the $c-v$ overlap integral,
\begin{multline}
 	\langle {\mathcal F}_e({\bf k}_i^{\mathrm{th}},\xi_i)|{\mathcal F}_{hh}({\bf k}_3^{\mathrm{th}},\xi_3) \rangle \approx\\
 		\langle {\mathcal F}_e^{(0)}(\mu{\bf k}_g,\xi_i)|{\mathcal F}_{hh}^{(2)}(-{\bf k}_g,\xi_3) \rangle,
 \label{eq_Icv1}
\end{multline}
where $|{\mathcal F}_e^{(0)}\rangle$ is a pure $s$-type state, corresponding to a single-band approximation for the final low-energy states having much smaller wave vector than that of the initial states (0 and 3). Therefore corrections to $|{\mathcal F}_e^{(0)}\rangle$ do not enter Eq.~(\ref{eq_Icv1}) in the leading order in $\mu=m_e/m_{hh}$.

\section{Approximate angular dependence of the quadratic term at arbitrary $\beta$}
\label{Sec_Ap:K_full}

Even thought an exact angular dependence of the quadratic term~(\ref{W2}) at arbitrary $\beta$ can be calculated only numerically, analytic approximation to it can be constructed via Pade-Borel method. For the cases of strong and weak spin-orbit coupling we have calculated series expansion of Eq.~(\ref{W2}) by $1/\beta \ll 1$ and $\beta\ll 1$ up to second and third orders, respectively. The resulting approximating function for $\mathcal{K}(\textbf{u},\beta)$ is
\begin{gather}
	\label{Kapprox}
	\mathcal{K}({\bf u},\beta)=\mathcal{K}_1({\bf u},\beta)+\mathcal{K}_2({\bf u},\beta)\\
	\mathcal{K}_1({\bf u},\beta)=\frac{K_1({\bf u})+\beta K_2({\bf u})+\beta^2 K_3({\bf u})}{S({\bf u},\beta)(\beta+S({\bf u},\beta))}\\
	\mathcal{K}_2({\bf u},\beta)=\frac{K_2({\bf u})+\beta \tilde{K}_2({\bf u})}{\beta+S({\bf u},\beta)}\\
	S({\bf u},\beta)=\sqrt{4\mathcal{I}^2-12\mathcal{J}+\beta^2}\\
	K_1({\bf u})=-4(\mathcal{I}^2-3\mathcal{J})(4\mathcal{I}^2-\mathcal{I}-3\mathcal{J})\\
	K_2({\bf u})=-8\mathcal{I}^3+2\mathcal{I}^2+18\mathcal{I}\mathcal{J}-4\mathcal{J}\\
	K_3({\bf u})=-2\mathcal{I}^2+\mathcal{I}-3\mathcal{J}\\
	\tilde{K}_2({\bf u})=3\mathcal{J}+\mathcal{I}-4\mathcal{I}^2.
	\label{KapproxLast}
\end{gather}

\bibliography{Bib_IIR_JAP}

\end{document}